\documentclass[aps,prl,twocolumn,groupedaddress,amsmath,superscriptaddress,amssymb,showpacs,10pt,reprint]{revtex4-1}
\usepackage{amsmath}
\usepackage{amsfonts}
\usepackage{amssymb}
\usepackage{graphicx}
\usepackage{cleveref}
\usepackage[none]{hyphenat}
\usepackage{mathptmx}
\usepackage{graphicx,graphics,color,epsfig}
\usepackage{amsfonts}
\usepackage{epstopdf}
\usepackage{bm}
\usepackage{times,xspace}
\usepackage{color}
\usepackage[utf8]{inputenc}

\def\beq{\begin{eqnarray}}\def\eeq{\end{eqnarray}}
\def\be{\begin{equation}}\def\ee{\end{equation}}

\begin{document}

\title{Direct Evidence for Time-Reversal Symmetry Breaking in Topological Superconductor $ \text{Sr}_{0.1}\text{Bi}_2\text{Se}_3 $ }
\author{P. Neha}
\affiliation{Jawaharlal Nehru University, New Delhi-110067, India}
\author{P.K. Biswas}
\affiliation{ISIS Pulsed Neutron and Muon Source,STFC Rutherford Appleton Laboratory, Harwell Campus,Didcot, Oxfordshire-OX110QX, United Kingdom}
\author{Tanmoy Das}
\email[corresponding author:]{tnmydas@gmail.com}
\affiliation{Department of Physics, Indian Institute of Science, Bangalore-560012, India}
\author{S. Patnaik}
\email[corresponding author:]{spatnaik@mail.jnu.ac.in}
\affiliation{Jawaharlal Nehru University, New Delhi-110067, India}

\begin{abstract}
 The single helical Fermi surface on the surface state of three-dimensional topological insulator Bi$_2$Se$_3$ is constrained by the time-reversal invariant bulk topology to possess a spin-singlet superconducting pairing symmetry. In fact, the Cu-doped, and pressure-tuned superconducting $\rm{Bi}_2\rm{Se}_3$ show no evidence of the time reversal symmetry breaking. We report on the detection of the time reversal symmetry (TRS) breaking in the topological superconductor $\rm{Sr}_{0.1}\rm{Bi}_2\rm{Se}_3$, probed by zero-field (ZF) $\mu$SR measurements. The TRS breaking provides strong evidence for the existence of spin-triplet pairing state. The temperature dependent super-fluid density deduced from transverse-field (TF) $\mu$SR measurement yields nodeless superconductivity with low superconducting carrier density and penetration depth $ \lambda $ = 1622(134) nm. From the microscopic theory of unconventional pairing, we find that such a fully gapped spin triplet pairing channel is promoted by the complex interplay between the structural hexagonal warping and higher order Dresselhaus spin-orbit coupling terms. Based on Ginzburg-Landau analysis, we delineate the mixing of singlet to triplet pairing symmetry as the chemical potential is tuned far above from the Dirac cone. Our observation of such spontaneous TRS breaking chiral superconductivity on a helical surface state, protected by the TRS invariant bulk topology, can open new avenues for interesting research and applications.
\end{abstract}

\maketitle
The nuances of superconducting state derived from topological insulators have attracted significant attention in the recent past and have provided a fertile testing ground for several emergent phenomena associated with quantum condensed matter\cite{topologicalinsulator}. Along with non-trivial bulk wave functions, topological superconductors are associated with a set of symmetry principles and are predicted to host unconventional superconductivity with additional symmetry breaking paradigm\cite{symmetry}. A topological insulator is characterized by an insulating bulk and gapless conducting surface states. In analogy, a topological superconductor is assigned to be fully gapped in the bulk along with gapless surface Andreev bound states\cite{zhang}. Such exotic Field- theoretic ideas have provided the material basis for the realization of Majorana fermions, the untraced elementary particle that is its own anti-particle, and their projected usage in quantum computers\cite{majoranacomp}. Experimentally, on the other hand, the onset of superconductivity in metal intercalated 3-Dimensional topological insulator $ \rm{Bi}_2\rm{Se}_3 $ has provided access to decipher pairing and order parameter symmetry in topological superconductors\cite{unconv}. However, evidence for the Andreev surface states in Cu, Sr, and Nb intercalated $\rm{Bi}_2\rm{Se}_3 $ superconductors has remained controversial\cite{andreev}. Additionally, photo-emission measurements have presented the evidence for an isolated Dirac cone near the Fermi level, without any intervening bulk states in the Sr$_{0.1}$Bi$_2$Se$_3$ samples\cite{arpes}. The linear dispersion of the surface state near the Dirac cone stems from the spin-momentum locking due to Rashba-type spin-orbit coupling (SOC). In a typical Rashba-type SOC in other 2D electron gases, the corresponding Fermi surface is split into two counter-helical pockets and mixing of singlet and triplet superconductivity with associated time reversal symmetry breaking (TRSB) is expected. However, owing to the $Z_2$  bulk topology, the surface states of Sr$_{0.1}$Bi$_2$Se$_3$ host only a single helical Fermi pocket. In such a Fermi surface topology, it is well known that only a singlet pair $({\bf k}\uparrow,-{\bf k}\downarrow)$ is allowed (${\bf k}$ is momentum and $\uparrow/\downarrow$ are spins). Moreover, since its spin flip partner, i.e, $({\bf k}\downarrow,-{\bf k}\uparrow)$ is absent in the single helical surface state, the antisymmetric requirement of the pair wavefunction prescribes that the order parameter must be odd-parity. The resulting order parameter therefore follows the underlying time-reversal symmetry\cite{LFuTSC}. 

Theoretically, materials having trigonal and hexagonal crystal structure with strong spin-orbit coupling are expected to be susceptible towards rotational or spin rotational symmetry breaking. In particular,  rotational symmetry breaking and consequent unconventional superconductivity in $\rm{Sr}_{0.1}\rm{Bi}_2\rm{Se}_3 $ and $\rm{Nb}_{0.25}\rm{Bi}_2\rm{Se}_3$ have been reported\cite{rotbrk}. The presence of topological order with Dirac type metallic surface states in superconducting topological insulators have also been detected \cite{arpes}. However, the experimental results on the pairing symmetry remain contradictory as both signatures for conventional and unconventional pairing symmetry are observed with different experiments\cite{pairingsymmetry}. Besides, a fascinating possibility of unconventional pairing mechanism arises with the development of local moment at Cooper pair sites because of the relative phase difference in a multicomponent order parameter. Such possibilities have recently been evidenced in $\rm{Sr}_{2}\rm{Ru}\rm{O}_{4} $ \cite{Luke} and in the case of non-centrosymmetric $\rm{Re}_6\rm{Zr}$,  associated TRS breaking is confirmed\cite{Singh}.

In this communication we summarise the results of Muon spin rotation and relaxation ($\mu$SR)measurements on the single crystals of topological superconductor $\rm{Sr}_{0.1}\rm{Bi}_2\rm{Se}_3 $. We provide strong evidence for coexisting time-reversal symmetry(TRS) broken states with triplet pairing along with TRS  invariant singlet pairing states. This is allowed by symmetry because of hexagonal wrapping effect and higher order spin orbit coupling effect which is specific to $\rm{Sr}_{0.1}\rm{Bi}_2\rm{Se}_3$.  Based on Ginzburg-Landau theory we also develop and specify the criterion for such mixed pairing state and sketch the phase diagram as a function of chemical potential in doped topological superconductors. 

Single crystals of $\rm{Sr}_{0.1}\rm{Bi}_2\rm{Se}_3 $ used in this study were grown by a modified Bridgeman technique. These single crystalline materials were extensively characterized by structural and magnetic measurements and a superconducting transition temperature ($ \text{T}_\text{c}$) of $\sim $ 2.5 K was ascertained\cite{srbise}. For the $\mu$SR experiments, about 3g of  powdered $\rm{Sr}_{0.1}\rm{Bi}_2\rm{Se}_3 $ crystals was mounted on a silver sample holder and placed in a dilution refrigerator operating in the temperature range of $0.05-5$ K. Muon spin rotation/relaxation measurements were carried out using the MuSR spectrometer\cite{isis} of the ISIS facility at the Rutherford Appleton Laboratory, UK and the measurements were performed under zero-field (ZF) and transverse-field (TF) protocols. In the ZF-$\mu$SR experiments, the sample was cooled down to 0.1 K in true zero-field condition to avoid trapping of any stray field and data were collected up to 3.7 K by warming the sample. Similarly,in the TF-$\mu$SR, the sample was first field cooled to the base temperature (0.09K) in a magnetic field of 10 mT applied (above $T_{\rm c}$) perpendicular to the initial muon spin polarization and $\mu$SR spectra were collected up to 3.6 K upon warming the sample. The ZF and TF-$\mu$SR data were analyzed using the software packages Mantid~\cite{mantid} and MUSRFIT\cite{Suter}.

\begin{figure}[ht]
\centering
\includegraphics[scale=0.3]{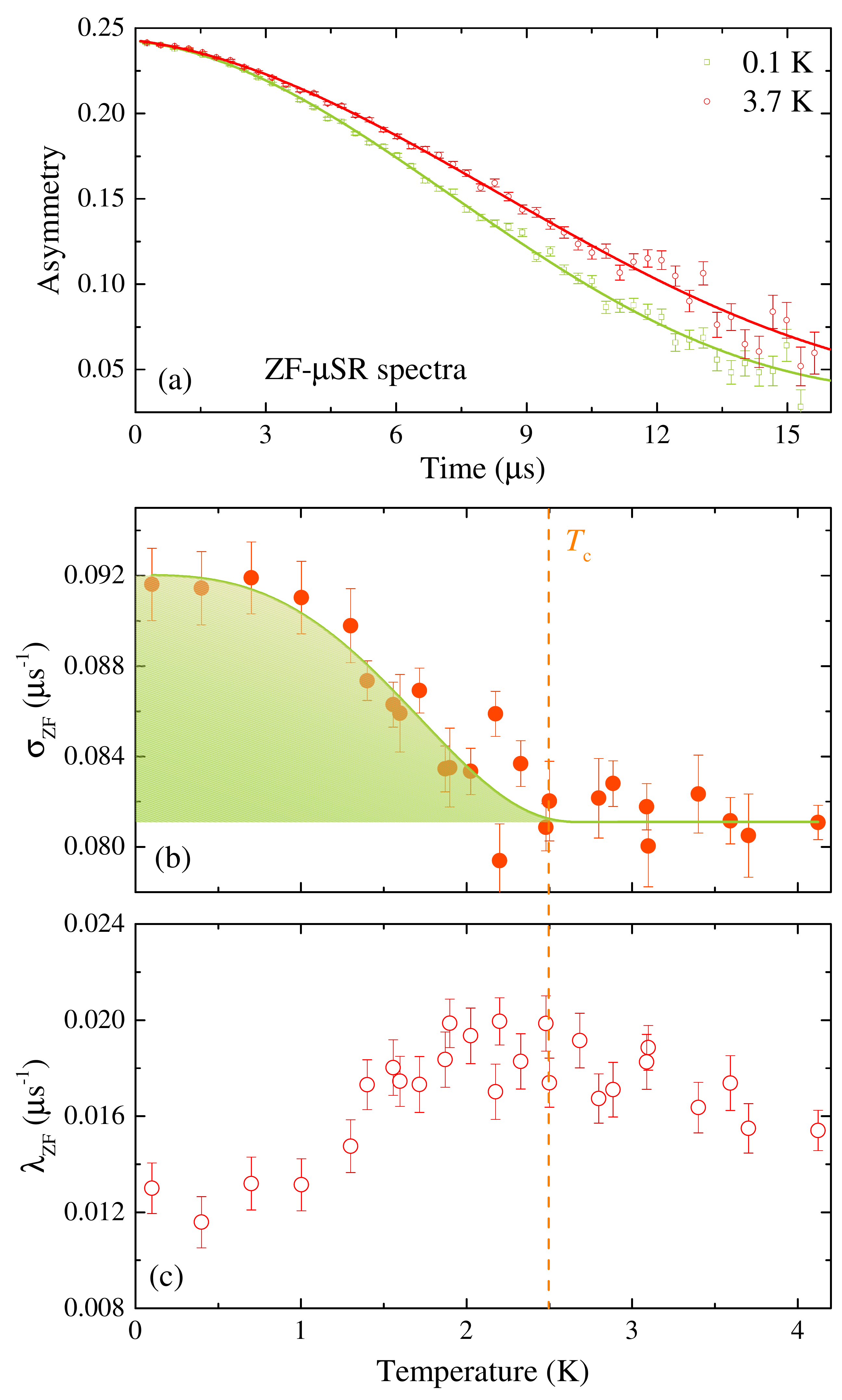}
\caption{\label{Fig1} (a) ZF-$\mu$SR spectra collected at 0.1K and 3.7 K. (b) Temperature variation of the muon spin relaxation rate $\sigma_{\rm{ZF}}$ above and below $T_{\rm c}$.(c) Variation in $ \lambda_{\rm{ZF}} $ with temperature also indicates existence of spontaneous field.}
 \label{fig:ZF}
\end{figure}

\section{Muon Spectroscopy}
The efficacy of the $\mu$SR technique towards unravelling the unconventional superconductivity rests on the fact that the magnetic moment associated with Cooper pairs in such cases is non zero which reflects the time reversal symmetry breaking \cite{unconvmuon}. The $\mu$SR technique in ZF mode is exceptionally sensitive to small changes in internal fields. It can easily measure local magnetic fields of the order of $\approx$ 10 $\mu$T that corresponds to about $ 10^{-2} $ of Bohr magnetron $\mu_{\rm B}$. ZF-$\mu$SR measurements are therefore useful to investigate any additional magnetic signal arising spontaneously with the onset of superconductivity. Fig.\ref{Fig1} shows the ZF-$\mu$SR spectra for $\rm{Sr}_{0.1}\rm{Bi}_2\rm{Se}_3 $ at (a) 0.1 K and (b) 3.7 K respectively. The Asymmetry parameter plotted in Fig.\ref{Fig1} is the time dependent normalized difference between number of positrons recorded in forward and backward detectors. ZF-$\mu$SR asymmetry spectra do not show any oscillatory signal which rules out the presence of large internal magnetic fields associated with long range magnetic order. It is evident that the data collected at 0.1 K (below $T_c $) shows higher relaxation than that at 3.7 K (above $T_c$). The ZF-$\mu$SR signal above $T_c$ shows small relaxation, arising from a temperature independent background contribution. The additional relaxation in the asymmetry signal below $T_c$ is due to small local fields arising in the superconducting state. Such static magnetic moments are associated with intrinsic Cooper pair magnetization. For a quantitative evaluation of the temperature dependence of the relaxation rate, ZF-$\mu$SR data were analysed using a static Gaussian Kubo-Toyabe relaxation function multiplied by an exponential relaxation function with decay rates $\sigma_{\rm{ZF}}$ and $\lambda_{\rm{ZF}}$ as

\begin{multline}
A(t)= A(0)\left\{\frac{1}{3}+\frac{2}{3}\left(1-{\sigma_{\text{ZF}}}^2t^2\right){\exp}\left(-\frac{{\sigma_{\text{ZF}}}^2t^2}{2}\right)\right\} \\
{\exp}(-\lambda_{\rm{ZF}})+A_{bg},
 \label{eq:KT_ZFequation}
\end{multline}
where $A(0)$, $A_{bg}$ are the initial and background asymmetries, and $\sigma_{\rm{ZF}}$ and $\lambda_{\rm{ZF}}$ are muon spin relaxation rates of the randomly orientated nuclear and electronic moments, respectively. The solid lines in Fig. \ref{fig:ZF} (a) are the fits to the data using the above equation.

In the fitting process, both relaxation rates $\sigma_{\rm{ZF}}$  and $\lambda_{\rm{ZF}}$ were set as free parameters in the first instance. Fig.\ref{fig:ZF} (b) shows the temperature dependence of $\sigma_{\rm{ZF}}$ which is displaying an increase in relaxation rate just below $T_{\rm c}$  of $\approx$ 2.5 K. This increase in relaxation rate $\sigma_{\rm{ZF}}$ reconfirms that spontaneous magnetic fields are emerging in the superconducting state of $\rm{Sr}_{0.1}\rm{Bi}_2\rm{Se}_3$. The appearance of such spontaneous fields in $\rm{Sr}_{0.1}\rm{Bi}_2\rm{Se}_3$ just below $T_{\rm c}$ provides strong evidence for a TRS breaking pairing state in this material. Similar behaviour has also been observed in the past by $\mu$SR in $\rm{Sr}_{2}\rm{Ru}\rm{O}_{4} $ \cite{Luke}, $\rm{Re}_6\rm{Zr}$ \cite{Singh},$ \rm{UPt}_3$ and $\rm{(U,Th)}\rm{Be}_{13}$ \cite{Luke1, Heffner},$ \rm{Pr}\rm{Os}_{4}\rm{Sb}_{12}$, $\rm{Pr}(\rm{Os}_{1-\rm{x}}\rm{Ru}_{\rm{x}})_{4}\rm{Sb}_{12} $, $\rm{Pr}\rm{Pt}_4\rm{Ge}_{12}$ \cite{Aoki, Shu, Maisuradze}, $\rm{La}\rm{Ni}\rm{Ga}_2$ \cite{Hillier} and SrPtAs \cite{Biswas}, etc. We also observe a broad hump in the temperature variation of $\lambda_{\rm ZF}$ (see Fig.\ref{fig:ZF} (c)). The broad hump in $\lambda_{\rm ZF}(T)$ is seen around $T_{\rm c}$ and is in agreement with the appearance of spontaneous magnetic moments. While a detailed description on the possible origins for the occurrence of TRS broken superconducting state in $\rm{Sr}_{0.1}\rm{Bi}_2\rm{Se}_3$ will be discussed later, in the following we discuss the issues related to symmetry of superconducting order parameters in topological superconductor $\rm{Sr}_{0.1}\rm{Bi}_2\rm{Se}_3 $. 

\begin{figure}[ht]
\centering
\includegraphics[scale=0.34]{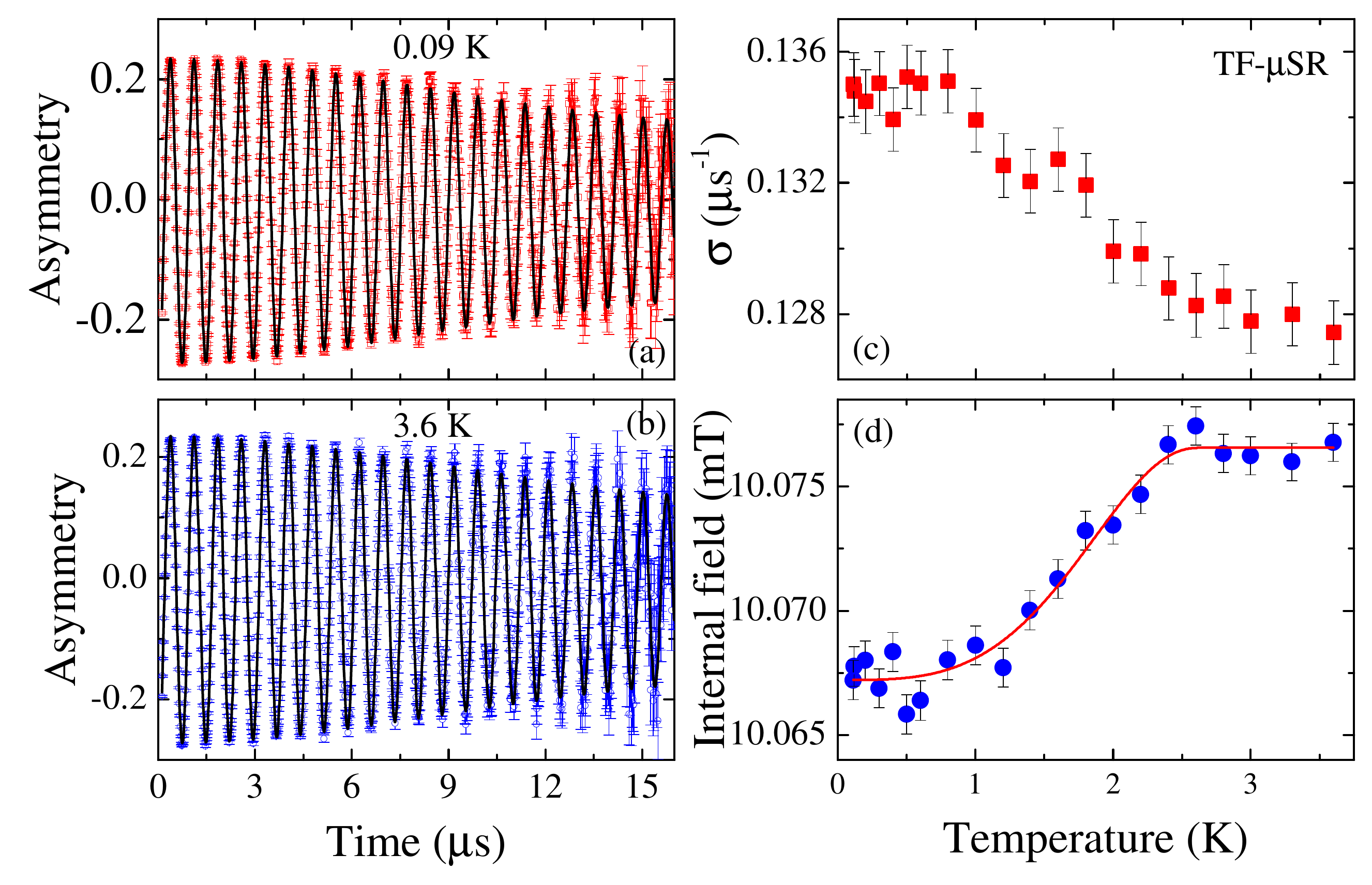}
\caption{\label{fig2}TF(Transverse Field) $\mu$SR spectra taken at (a) 0.09K and (b) 3.6K at 10 mT applied magnetic field (c) Variation of depolarization rate with temperature. (d) Variation of internal field in the material with temperature.}
 \label{fig:TF}
\end{figure}

In order to understand temperature dependence of super fluid density and the pairing symmetry of $\rm{Sr}_{0.1}\rm{Bi}_{2}\rm{Se}_{3}$, TF-$\mu$SR experiments were carried out in the superconducting mixed state under an applied magnetic field of 10 mT. Fig.\ref{fig:TF} (a) and (b) show the TF-$\mu$SR asymmetry spectra for $\rm{Sr}_{0.1}\rm{Bi}_{2}\rm{Se}_{3}$ collected at 0.09 K and 3.6 K, respectively. Data collected at 0.09 K shows higher relaxation compare to the normal state at 3.6 K due to inhomogeneous field distribution of flux line lattice. The solid lines in Fig. \ref{fig:TF} (a) and (b) are the fits to the data using a simple Gaussian type oscillatory distribution function

\begin{multline}
\label{Depolarization_Fit}
A^{TF}(t)=A(0){\exp}\left(\frac{-\sigma^{2}t^{2}}{2}\right)\cos\left(\gamma_\mu B_{\rm int} t +\phi\right) \\
+A_{\rm bg}(0)\cos\left(\gamma_\mu B_{\rm bg}t +\phi\right),
\end{multline}
where $A(0)$ and $A_{\rm bg}$(0) are the initial asymmetries of the sample and background signals, $\gamma_{\mu}/2\pi=135.5$~MHz/T is the muon gyromagnetic ratio~\cite{Sonier}, $B_{\rm int}$ and $B_{\rm bg}$ are the internal and background magnetic fields, $\phi$ is the initial phase of the muon precession signal, and $\sigma$ is the Gaussian muon spin relaxation rate.

The formation of flux line lattice is evident from the enhancement of relaxation rate $\sigma $ observed below transition temperature 2.5 K(Fig.\ref{fig:TF}(c)). The temperature dependence of the internal field shown in Fig. \ref{fig:TF}(d) displays a diamagnetic shift in the field distribution just below $T_{\rm c}$, a clear sign of bulk superconductivity in  this material. The total sample relaxation rate $\sigma$ comes from two contributions, the superconducting part $\sigma_{sc}$ due to formation of vortex lattice and the non-superconducting part $\sigma_{nm}$ due to the presence of nuclear dipole moments in the material. The later one is expected to be constant over the temperature range of this study. The superconducting component of the relaxation is obtained by quadratically subtracting the background nuclear dipolar relaxation rate obtained from TF-$\mu$SR spectra above $T_{c}$ as $\sigma=\sqrt{\sigma^{2}_{\rm sc} + \sigma^{2}_{\rm nm}}$.

\begin{figure}[ht]
\centering
\includegraphics[scale=0.35]{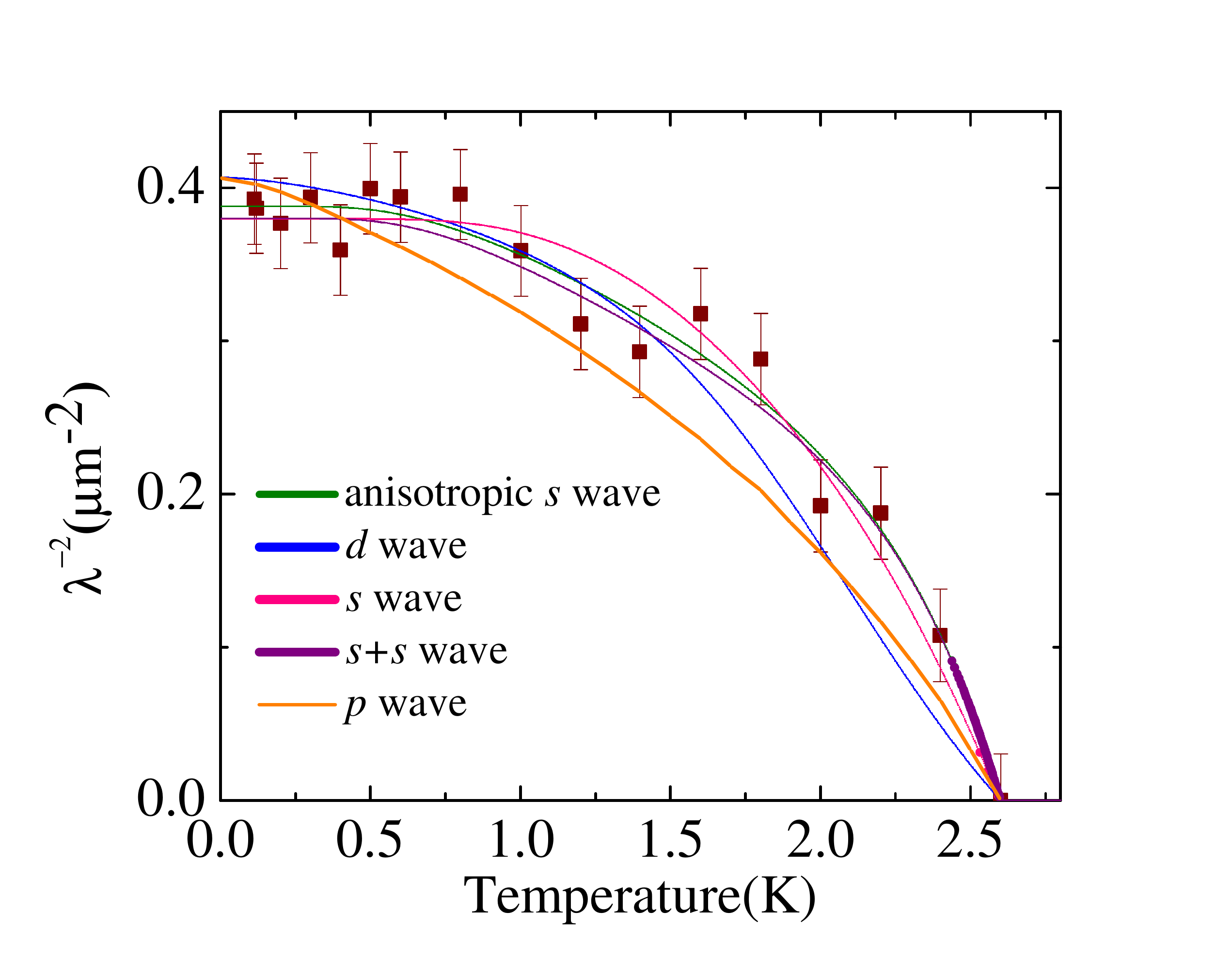}
\caption{Variation of ${\lambda}^{-2}$ as a function of temperature for $\rm{Sr}_{0.2}\rm{Bi}_{2}\rm{Se}_{3}$. Solid lines are the fit to the data using different models described in the text.}
 \label{fig:lambda}
\end{figure}

The temperature dependence of the magnetic penetration depth $\lambda(T)$ can be reconstructed from $\sigma_{sc}(T)$ by the simplified Brandt equation~\cite{Brandt},

\begin{equation}
\frac{\sigma_{sc}\left(T\right)}{\gamma_\mu}=0.06091\frac{\Phi_0}{\lambda^{2}\left(T\right)},
\end{equation}
where $\Phi_0=2.068\times10^{-15}$~Wb is the flux quantum. $\lambda^{-2}(T)$ is proportional to the effective superfluid density, $\rho_s \propto \lambda^{-2} \propto n_{\rm s}/m^*$ ($n_{\rm s}$ is the charge carrier concentration, and $m^*$ is the effective mass of the charge carriers) and hence bear the signature of the symmetry of the superconducting gap. Fig.~\ref{fig:lambda} shows the temperature dependence of $\lambda^{-2}$ and hence the superfluid density for $\rm{Sr}_{0.1}\rm{Bi}_{2}\rm{Se}_{3}$. The superfluid density shows saturation below ${T_{\rm c}}/3$ which in turn suggest the absence of low-lying excitations close to zero temperature that indicates nodeless superconductivity in $\rm{Sr}_{0.1}\rm{Bi}_{2}\rm{Se}_{3}$. To get a quantitative estimate, the $\lambda^{-2}(T)$ data were fitted using a single-gap $s$, $d$ and anisotropic $s$ wave or two-gap \textit{s+s} wave models using the following functional form~\cite{Carrington,Padamsee}:

\begin{equation}
\label{two_gap}
\frac{\lambda^{-2}\left(T\right)}{\lambda^{-2}\left(0\right)}=\omega\frac{\lambda^{-2}\left(T, \Delta_{0,1}\right)}{\lambda^{-2}\left(0, \Delta_{0,1}\right)}+(1-\omega)\frac{\lambda^{-2}\left(T, \Delta_{0,2}\right)}{\lambda^{-2}\left(0, \Delta_{0,2}\right)},
\end{equation}
where $\lambda\left(0\right)$ is the value of the penetration depth at $T=0$~K, $\Delta_{0,i}$ is the value of the $i$-th ($i=1$ or 2) superconducting gap at $T=0$~K and $\omega$ is the weighting factor of the first gap. Each term in Eq.~\ref{two_gap} is evaluated using the standard expression within the local London approximation ($\lambda \gg \xi$)~\cite{Tinkham, Prozorov} as

\begin{equation}
\frac{\lambda^{-2}\left(T, \Delta_{0,i}\right)}{\lambda^{-2}\left(0, \Delta_{0,i}\right)}=1+\frac{1}{\pi}\int^{2\pi}_{0}\int^{\infty}_{\Delta_{\left(T,\varphi\right)}}\left(\frac{\partial f}{\partial E}\right)\frac{ EdE d\varphi}{\sqrt{E^2-\Delta_i\left(T,\varphi\right)^2}},
\end{equation}
where $f=\left[1+\exp\left(E/k_{\rm B}T\right)\right]^{-1}$ is the Fermi function, $\varphi$ is the angle along the Fermi surface, and $\Delta_i\left(T,\varphi\right)=\Delta_{0, i}\delta\left(T/T_c\right)g\left(\varphi\right)$, where $g\left(\varphi\right)$ describes the angular dependence of the gap and it is 1 for $s$  and $s+s$ wave gaps, $\left|\sin\left(\varphi/2\right)\right|$ for $p$ wave gap, $\left|\cos\left(2\varphi\right)\right|$ for $d$ wave gap and $\left(s+\cos4\varphi\right)$ for anisotropic $s$ wave gap. An approximation to $\Delta(T)$ can be written as $\delta\left(T/T_{\rm c}\right)=\tanh\left\{1.82\left[1.018\left(T_{\rm c}/T-1\right)\right]^{0.51}\right\}$~\cite{Carrington}.

\begin{table}
\caption{Fitted parameters to the $\lambda^{-2}(T)$ data of $\rm{Sr}_{0.1}\rm{Bi}_{2}\rm{Se}_{3}$ using the different models as described in the text.}
\label{table_of_gapratios}
\begin{center}
\begin{tabular}[t]{lll}\hline\hline
Model & Gap value (meV) & $\chi ^2$\\\hline
$s$-wave & $\Delta$=0.49(4) & 1.40\\
$s+s$-wave & $\Delta_1$=0.7(3), $\Delta_2$=0.3(1) and $\omega=0.58(8)$ & 1.06\\
anisotropic $s$-wave & $\Delta=0.54(6)$ with $s=0.6(2)$ & 1.02\\
$p$-wave & $\Delta$=0.55(3) & 4.27\\
$d$-wave & $\Delta$=0.4(1) & 3.53\\\hline\hline
\end{tabular}
\par\medskip\footnotesize
\end{center}
\end{table} 

The solid curves shown in Fig.~\ref{fig:lambda} are the fits to the $\lambda^{-2}(T)$ data using different gap models. All the fitted parameters are summarized in Table~\ref{table_of_gapratios}. Both anisotropic $s$ wave and two-gap $s+s$ wave gap models give the lowest $\chi_{\rm reduced}^2$ value and hence give the best fit to the data compared to any other models mentioned above. Recent scanning tunnelling microscopy measurements shows two-gap \textit{s+s} wave and anisotropic \textit{s} wave pairing symmetry as the prominent pairing mechanism\cite{stm}. For single-gap nodeless pairing we estimate $\Delta$=0.49(4) meV. The value of the gap to $T_{\rm c}$ ratio, $\Delta/\kappa_{\rm B}T_{\rm c}$ is 2.18 which is significantly higher than the BCS value of 1.76. Given the roughness in the data points, it is difficult to pin point the exact symmetry of the gap structure in $\rm{Sr}_{0.1}\rm{Bi}_{2}\rm{Se}_{3}$. The possibility of ${\rm {p}_{\rm {x}}}+i{\rm{p}_{\rm {y}}}$ pairing symmetry with two gaps one corresponding to singlet and another corresponding to triplet pairing symmetry is also not ruled out. However, within the present statistical accuracy it is implied that $\rm{Sr}_{0.1}\rm{Bi}_{2}\rm{Se}_{3}$ may not have the extended nodes in the gap function. The absolute value of the magnetic penetration depth $\lambda(0)$ is calculated to be 1622(134) nm, which is in close agreement with the previous documented experimental value \cite{srbise}. The remarkably large value of $\lambda(0)$ indicates the presence of very low superconducting carrier density, a common feature in this class of topological superconductors.\\

In summary therefore, while our TF-$\mu$SR data that probe temperature dependent penetration depth around vortices reflect nodeless superconductivity, the ZF-$\mu$SR results that probe small moments associated with Cooper pairs reflect odd parity pairing. Such singlet-triplet mixing is common in superconductor with additional symmetry breaking such as non-centrosymmetric superconductors. In the following, we develop a phenomenological theory for this observation with regard to topological superconductivity. 

\begin{figure}[ht]
\centering
\includegraphics[scale=0.6]{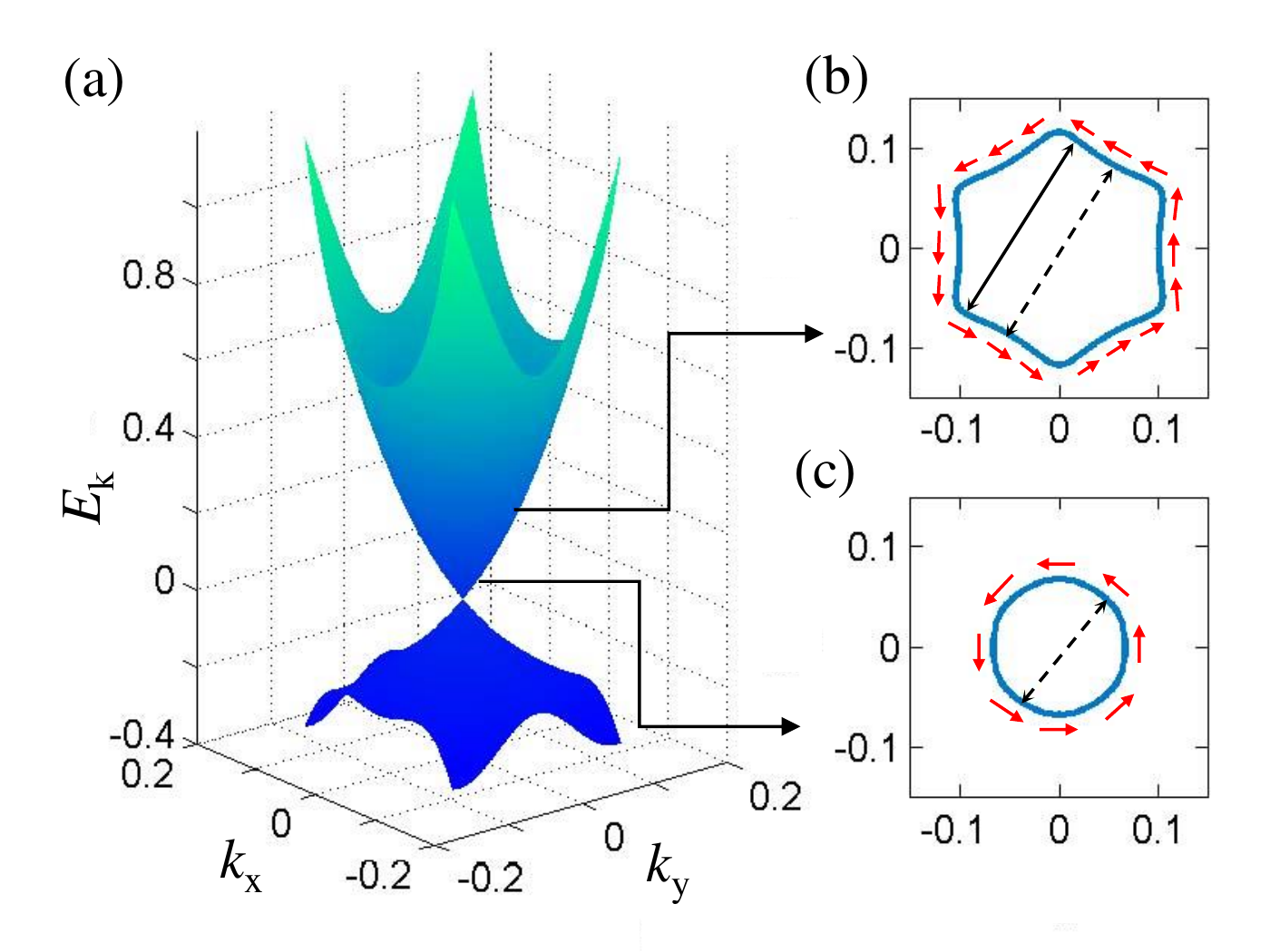}
\caption{\label{fig4} (a) Single helical surface state of a 3D topological insulator in the $k_x-k_y$-plane. (b-c) Constant energy cuts (b) far from the Dirac cone and (c) close to the Dirac cone. Red arrows show the corresponding spin alignments across the corresponding constant energy contours. Dashed arrow dictated the $k$ and $-k$ points in which spin is anti-parallel, while the solid arrows shows similar points where anti-parallel alignment of the spin is weakened due to higher order SOC and hexagonal warping effects.}
\end{figure}
\section{Theory}

In several superconductors coexisting singlet and triplet pairing mechanism with TRSB has been observed  before\cite{Singh,Suter,Luke1,Heffner,Aoki,Hillier}. However, the surface states of Sr$_{0.1}$Bi$_2$Se$_3$ can only host a single helical Fermi pocket that would result an odd parity TR invariant superconductivity. Therefore, the  experimental observation of the TRS breaking pairing demands a theoretical explanation which is beyond the existing understandings. 

In Sr$_{0.1}$Bi$_2$Se$_3$, the chemical potential lies far above the Dirac cone. It is well established that the spin-momentum locking weakens as we move away from the Fermi level, due to the structural hexagonal warping effects,\cite{LFuWarping}, as well as from higher order SOC effects.\cite{Basak} As a combined effect, as one moves away from the Dirac cone, the Fermi surface deviates from circular to snowflake type (see Fig.~\ref{fig4}). The corresponding spin texture becomes more anisotropic, as measured by photo-emission spectroscopy,\cite{warpedspin} This means, the spin of the electrons at a momentum $\pm{\bf k}$ away from the diagonal directions (shown in solid arrow in Fig.~\ref{fig4}(b)) are no longer anti-parallel to each other. Consequently, both TR-invariant singlet pairing (across dashed arrow) and TR-breaking triplet pairing (across the solid arrow) channels become allowed by symmetry. 

The warped surface state and its unconventional spin-texture can be realistically modelled by a low-energy Hamiltonian\cite{Basak} $H({\bf k})=\xi_{\bf k}\mathbf{I}_{2\times 2} + {\bf d}_{\bf k}\cdot {\bf \sigma}$, with ${\bf \sigma}$ being the $2\times 2$ Pauli matrices. The on-site dispersion $\xi_{\bf k}=k^2/m_1 + k^4/m_2$, with $k=|{\bf k}|$. The off-diagonal gap terms are $d_{x}=-\alpha_{\bf k} k_{y} -{\rm Im}[\beta_{{\bf k}}]$, and $d_{y}=\alpha_{\bf k} k_{x} -{\rm Re}[\beta_{{\bf k}}]$, where $\alpha_{\bf k}=\alpha_0+\alpha_1k^2 + \alpha_2 k^4$ and $\beta_{\bf k}=\beta_0[(k_+^5 + k_-^5) + i(k_{+}^5 - k_{-}^5)]$ are the first and fifth order Dresselhaus SOC coefficients, where $k_{\pm}=k_x\pm i k_y$. The structural warping term gives an Ising like, anisotropic spin-splitting as $d_z({\bf k}) = \lambda_{\bf k}(k_{+}^3 + k_{-}^3)$, with $\lambda_{\bf k}=\lambda_0+\lambda_1 k^2$. All parameters are obtained by fitting to the experimental Fermi surface warping and the anomalous spin-texture as given in Ref.~\cite{TBparameter,Basak,LFuWarping} Clearly, the SOC term $d_{||}=\sqrt{d_x^2 + d_y^2}$ provides the helicity to the electron's spin-momentum relationship, while the warping term $d_z$ opposes it. 
Clearly, $d_{||}$ promotes the spin-singlet pairing, while $d_z$  helps stabilizing the spin-triplet pairing.

\begin{figure}[ht]
\centering
\includegraphics[scale=0.51]{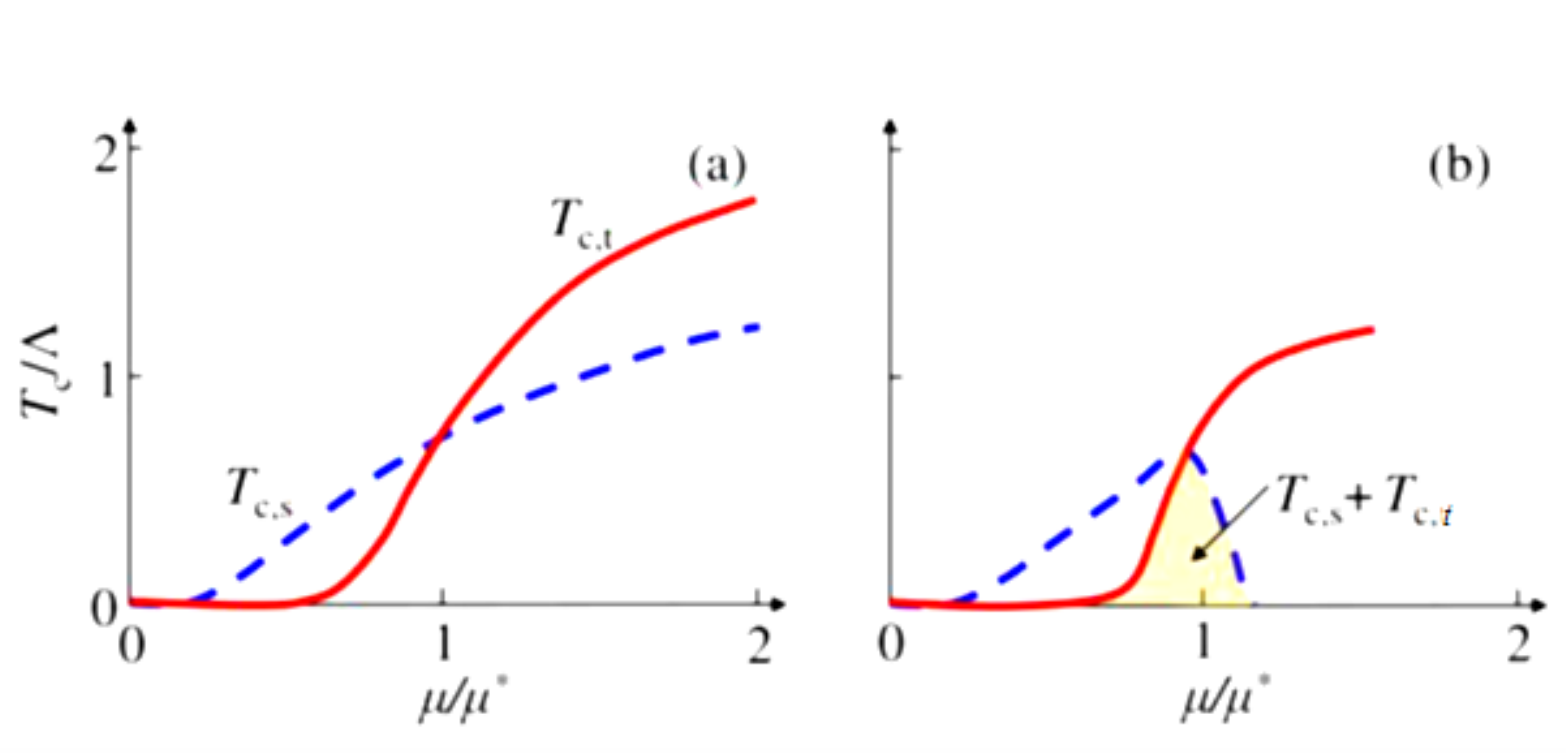}
\caption{\label{fig5} {(a) Two superconducting transition temperature for the singlet ($T_{c,s}$) and triplet ($T_{c,t}$) pairings, assuming same coupling constant $u_s=u_t$ but different density of states. The cross terms in the Free energy ($\gamma$, $\delta$) are neglected. (b) As the cross terms are introduced, both phases become complementary, with a small region of their coexistence near the tri-critical point $\mu^*$. }}
\end{figure}

The superconductivity is expected to have odd-parity symmetry, so it makes sense to study its unconventional pairing mechanism. Since, the spins are locked to the in-plane momentum, the relevant interaction term consists of Hubbard interaction and/or XY-type Heisenberg term for the spin-singlet pairing, and a Dyzoloshinskii-Moriya (DM) term for the spin-triplet pairing. The net interaction term is
\beq
%
H_{\rm int} &=& \sum_{i\ne j}\left[Un_{i\uparrow}n_{i,\downarrow} + J{\bf S}_{i}\cdot {\bf S}_{j} + i{\bf D}\cdot \left({\bf S}_{i}\times {\bf S}_{j}\right)\right],\nonumber\\
\label{Hintsupp1}
\eeq
where $U$ is the Hubbard interaction, $J$, and ${\bf D}$ are the nearest neighbor symmetric, and antisymmetric (DM) spin exchange terms, respectively. $n_{i,\sigma}$, and ${\bf S}_i$ are the number and spin density operators, respectively at a given site $i$, with spin $\sigma=\uparrow,\downarrow$. Given that spin is confined only in the $x,y$ plane, we set ${\bf D}=D_z\hat{z}$. Expanding the density, and spin operators in terms of fermionic creation and annihilation operators $c^{\dag}_{{\bf k},\sigma}$, $c_{{\bf k},\sigma}$, and using Hubbard-Stratonovic transformation,  we obtain  
the singlet and triplet SC order parameters,\cite{SM} defined as
\beq
\Delta_{s}({\bf k}) &=& \sum_{{\bf k}'}u_{s}({\bf k},{\bf k}')\left\langle c_{-{\bf k}',\downarrow} c_{{\bf k}',\uparrow}\right\rangle,\\
\Delta_{t}({\bf k}) &=& \sum_{{\bf k}'}u_{t}({\bf k},{\bf k}')\left\langle c_{-{\bf k}',\uparrow}c_{{\bf k}',\uparrow} \right\rangle.
\eeq
$c_{{\bf k},\sigma}$ is the annihilation operator of electron at momentum ${\bf k}$, with spin $\sigma$. The singlet and triplet pairing potential can be easily read as $u_{s}({\bf k},{\bf k}')=(Us_{{\bf k}}-Js_{-{\bf k}})s_{{\bf k}'}$, and $u_{t}({\bf k},{\bf k}')= D_z t_{\bf k}t_{{\bf k}'}$, where $s_{\bf k}$ and $t_{\bf k}$ are the structure factors for the singlet and triplet pairings. As discussed before, both $s_{\bf k}$ and $t_{\bf k}$ must be odd under parity. It is easy to verify that that due to the odd-parity nature of the pairing symmetry, $\Delta_s$ is TR invariant, while $\Delta_t$ breaks this symmetry. We note that owing to rhombohedral structure of this compound, the irreducible representation is reduced from six-fold symmetry to the three-fold $C_{3v}$ class. This also reflects in the SC order parameter, and can be a candidate explanation to the in-plane rotational symmetry breaking as observed before.\cite{rotbrk}

With an eye on the experimental observation, we are here mainly interested in unraveling the phase diagram between the $\Delta_s$, and $\Delta_t$ order parameters as a function of chemical potential tuning. The phase competition between the two order parameters can be understood within the Ginzburg-Landau framework, in which the Free energy can be expanded in terms of the order parameters as \cite{SM} 
\beq
F&=&\alpha_s|\Delta_s|^2 + \alpha_t|\Delta_t|^2 +\frac{\beta_s}{2}|\Delta_t|^4 + \frac{\beta_t}{2}|\Delta_t|^4 \nonumber\\
&&+ \frac{\gamma_s}{2}(\Delta_s\Delta^*_t)^2 + \frac{\gamma_t}{2}(\Delta_s^*\Delta_t)^2+\delta|\Delta_s|^2|\Delta_t|^2 .
\eeq
$i=s,t$ are for singlet and triplet terms, respectively. The full expressions for the expansion parameters $\alpha_i$, $\beta_i$, $\gamma$, and $\delta$ are given in the supplementary materials\cite{SM}. In the absence of phase competition terms, i.e., when $\gamma=\delta=0$, the individual phase transition of $\Delta_i$ occurs at $\alpha_i=0$.  Near the phase transition, $\alpha_i$ can be evaluated analytically, as $\alpha_i = \frac{1}{u_i} - N_i\log{\frac{\Lambda}{T}}$ ($i$=s,t), where $\Lambda$ is the momentum cutoff.  In the leading order terms, we have $N_s\sim\langle d_{||}({\bf k})\rangle_{\rm FS}$, and $N_t\sim\langle d_{z}({\bf k})\rangle_{\rm FS}$. In what follows, $N_s$ is determined by the SOC term $d_{||}$, while $N_t$ depends on the hexagonal warping term $d_z$. So, $N_s$ dominates near the Dirac cone, while $N_t$ takes over at higher energy. The individual SC transition temperature is $T_{c,i}=\Lambda e^{-1/u_iN_i}$, and its variation with the chemical potential is plotted in Figure\ref{fig5}(a). From scaling analysis, we can estimate that such transition occurs when the chemical potential becomes $\mu^*\sim \alpha_0 k_F$. 

As the phase competition terms $\gamma>0$ and $\delta>0$ are turned on, the phase diagram changes as follows. Once again the calculation simplifies at the critical point where $T_c^*=T_{c,s}=T_{c,t}$. At this point, $u_sN_s\approx u_tN_t$ which leads to $N_s/N_t\sim D_z/(U-J)$ at $\mu^*$. In this limit, we also find that $\beta_s=\beta_t=5\beta$, and $\gamma_s=\gamma_t=\delta=3\beta$, with $\beta=7\zeta(3)/64\pi^2(T_c^*)^2$. The Free energy is minimum when the $(\Delta_s\Delta_t^*)^2=-|\Delta_s|^2|\Delta_t|^2$, implying that the phase difference between the two order parameters is $\pi/2$. The Free energy minimization leads to the condition that both phases coexist when $\gamma\delta<\beta_s\beta_t$.\cite{GLtwophases} Since this condition is satisfied near $T_c^*$, we conclude that the singlet to triplet phase transition is intervened by a region of their uniform coexistence. Based on these results, we draw the over phase diagram as shown in Figure\ref{fig5}(b).\\

In summary, we have performed the ZF and TF $\mu$SR measurements on topological superconductor $ \rm{Sr}_{0.1}\rm{Bi}_{2}\rm{Se}_{3} $. The $\mu$SR measurements in ZF mode unveils the presence of triplet pairing with unambiguous evidence for time reversal symmetry breaking. The TF $\mu$SR measurement, on the other hand yields the presence of low carrier density, nodeless superconductivity and the zero temperature penetration depth is estimated to be $ \lambda(0) $ = 1622(134)nm. Theoretically, the existence  of triplet pairing is defined in terms of hexagonal wrapping effect with higher order Dresselhaus SOC terms. Under the framework of Ginburg-Landau theory, coexistence of singlet and triplet pairing is indicated                                                                                                                                                                                                                                                                    in terms of chemical potential tuning. Our observation of time reversal symmetry broken states in a novel class of topological superconductors is a surprising development that promises new insight into  superconducting states derived from topological insulators.

\section{Method}
{\bf Sample Synthesis:} Single crystals of $\text{Sr}_{0.1}\text{Bi}_2\text{Se}_3 $ were prepared by modified Bridgeman technique\cite{srbise}.  High-purity constituent  elements Bi, Sr and Se were taken in stoichiometry ratio in a quartz ampoule which was sealed under vacuum ($ 10^{-4}$ mbar) and then heated at $ 850^{0}$ for 8 days followed by slow cooling down to $650^{0}$ at the rate of $10^{0}$C/h. The ampoule was then quenched in ice cold water.

{\bf{$\mu$}{\bf SR}}{\bf Technique:} From the experimental technique point of view, $\mu$SR spectroscopy is established to be the most appropriate tool to decipher such ultra low magnetic moments associated with Cooper pairs. 
In a typical $\mu$SR experiment, 100 \% spin polarised muons are embedded onto the sample where they decay within 2.2$\mu$s giving rise to positrons. The  muon spin precesses at the local magnetic environment and the resultant positrons carry that information into detectors placed in the forward and backward direction of the muon beam.In essence, $\mu$SR spectroscopy can reflect signatures of unconventional superconductivity, time reversal symmetry breaking, coexistence of superconductivity and magnetism along with superconducting pairing symmetry. The Asymmetry parameter plotted here is the time dependent normalized difference function $\left[A(t)=\dfrac{N_{F}(t)-\alpha N_{B}(t)}{N_{F}(t)+\alpha N_{B}(t)}\right]$ between the number of positrons recorded in the forward detector $\rm{N}_{\rm{F}}(\rm{t}) $ and backward detector $\rm{N}_{\rm{B}}(\rm{t}) $ . $\alpha$ is the calibration constant which is determined by applying a transverse-field (TF) of 20 mT.

\section{Acknowledgements}
PN thanks University Grant Commission (UGC) for providing Basic Science Research(BSR) fellowship. SP acknowledges DST for supporting low temperature measurement facility at JNU (EMR/2016/003998/PHY). PN and SP thank Rutherford Appleton Laboratory for providing Newton Bhabha funding for the Muon spectroscopy measurements. TD acknowledges the financial support from the DST, India under the Start Up Research Grant (Young Scientist) [SERB No: YSS/2015/001286], and from Infosys young investigator award.


\begin{thebibliography}{35}
\bibitem{TBparameter} The values of the parameters are obtained by fitting the Fermi surface as well as the spin texture of Bi$_2$Te$_3$ as obtained by ARPES measurements.\cite{Basak} The parameters are $m_1=70.42$ eV$^{-1}${\AA}$^{-2}$, $m_2=76.92$ eV$^{-1}$\AA$^{-4}$, $\alpha_0=-5\times 10^{-4}$ eV\AA, $\alpha_1=-74\times 10^{-4}$ eV\AA$^3$, $\alpha_2=6.3\times 10^{-4}$ eV\AA$^5$, $\lambda_0 = -0.04$ eV\AA$^3$, and $\lambda_1 = 4\times10^{-5}$ eV\AA$^5$, and $\zeta=0.35$ eV\AA$^5$.
\end{thebibliography}
\end{document}